\documentstyle[11pt]{article}
\input epsf
\begin{document}
\thispagestyle{empty}
\begin{center}
\null\vspace{-1cm} {\footnotesize
{\tt }}\hfill VACBT/05/12, GNPHE/05/12 and IC/2005\\
\vspace{1cm}
\medskip
\vspace{2.5 cm} {\large \textbf{SOME ASPECTS OF MOYAL DEFORMED INTEGRABLE SYSTEMS}}\\
\vspace{2 cm} \textbf{O. DAFOUNANSOU}, $^{a}$\footnote{Junior
Associate of the Abdus Salam ICTP} \textbf{A. EL BOUKILI} $^{b}$
and \textbf{M.B. SEDRA}$^{c}$ \footnote{Corresponding author,
Senior
Associate of the Abdus Salam ICTP: sedra@ictp.it} \\
$^{a,\,c}${\small \ International Centre for Theoretical Physics,
Trieste, Italy.}\\
$^{a,\,b,\,c}$ {\small \ Virtual African Center For Basic Sciences
and Technology, VACBT,\\ Focal point: Lab/UFR-Physique des Haute
Energies, Facult´e des Sciences,
Rabat, Morocc}.\\
$^{b,\,c}$ {\small \ Groupement National de Physique de Hautes
Energies, GNPHE, Rabat, Morocco,}\\{\small \ \ \ }$^{a}${\small \
\ University of Douala, Faculty of
Sciences, Department of Physics, Douala. Cameroon,}\\
{\small \ \ \ } $^{b,\,c}${\small \ Universit\'{e} Ibn Tofail,
Facult\'{e} des
Sciences, D\'{e}partement de Physique,}\\
{\small \ \ \ }{\small \ Laboratoire de Physique de la Mati\`{e}re
et Rayonnement (LPMR), K\'{e}nitra, Morocco}\\

\end{center}
\vspace{0.5cm}
\centerline{\bf Abstract}
\baselineskip=18pt
\bigskip
Besides its various applications in string and D-brane physics,
the $\theta$-deformation of space (-time) coordinates (naively
called the noncommutativity of coordinates), based on the
$\star$-product, behaves as a more general framework providing
more mathematical and physical informations about the associated
system. Similarly to the Gelfand-Dickey framework of pseudo
differential operators, the Moyal $\theta$-deformation applied to
physical problems makes the study more systematic. Using these
facts as well as the backgrounds of Moyal momentum algebra
introduced in previous works \cite{ref 21, ref 25, ref 26}, we
look for the important task of studying integrability in the
$\theta$-deformation framework. The main focus is on the
$\theta$-deformation version of the Lax representation of two
principal examples: the $sl_2$ KdV$_{\theta}$ equation and the
Moyal $\theta$-version of the Burgers systems. Important
properties are presented.\\\\


\hoffset=-1cm \textwidth=11,5cm \vspace*{1cm}
\hoffset=-1cm\textwidth=11,5cm \vspace*{0.5cm}

\newpage
\section{Introduction}
It's known for several years that $(1+1)$-dimensional integrable
models \cite{ref 1} in connection with conformal field theories
\cite{ref 2} and their underlying lower $(s\leq 2)$ \cite{ref
3,ref 4,ref 5,ref 6,ref 7} and higher $(s\geq 2)$ \cite{ref 8, ref
9} spin symmetries, play a central role in various areas of
research. In this sense the Virasoro algebra $W_2$ defines the
second Hamiltonian structure for the KdV hierarchy \cite{ref 10},
$W_3$ rely to the Boussinesq \cite{ref 11} and
$W_{1+\infty}$ is associated to the KP hierarchy \cite{ref 12} and so on.\\\\
Note by the way that the methods developed for the analysis of
integrable models can be also used to study the well known problem
related to hyper-Kahler metrics building program. This is an
important question of hyper-Kahler geometry, which can be solved
in a nice way in the harmonic superspace, see for instance
\cite{ref 13}.\\\\
In the Gelfand-Dickey framework of integrable systems, one usually
introduce the following operators \cite{ref 14}
\begin{equation}
{\mathcal L}_{n}=\sum_{j\in Z}u_{n-j}\partial^{j},
\end{equation}
These pseudo-differential Lax operators, allowing both positive as
well as nonlocal powers of the differential $\partial ^{j}$, are
those used to establish the correspondence between KdV hierarchies
and extended conformal symmetries. The fields $u_s$ of conformal
spin $s$ can be related to general primary fields upon some
covariantization procedure \cite{ref 15}.\\\\
During the last several years, there has been a growth in the
interest in non-commutative geometry (NCG), which appears in
string theory in several ways \cite{ref 16}. Much attention has
been paid also to field theories on noncommutative (NC) spaces and
more specifically Moyal deformed space-time, because of the
appearance of such theories as certain limits of string, D-brane
and M-theory \cite{ref 17}. Non-commutative field theories
emerging from string (membrane) theory  stimulate actually a lot
of important questions about the non-commutative integrable
systems and how they can be described in terms of star product and
Moyal bracket \cite{ref 18, ref 19, ref 20, ref 21, ref 22, ref
23, ref 24, ref 25, ref 26,ref 27}.\\\\
Recall that in the Moyal momentum algebra\footnote{The Moyal
momentum algebra, as a notion, was introduced first by the authors
of \cite{ref 21, ref 22}, and systematically studied later by the
authors of \cite{ref 25, ref 26} with some application to
conformal field theory and $\theta$-deformed integrable models} as
a the ordinary pseudo differential Lax operators (1) are naturally
replaced by momentum Lax operators
 \begin{equation}
{\mathcal L}_{n}=\sum_{j\in Z}u_{n-j}\star p^{j},
\end{equation}
satisfying a non-commutative but associative algebra inherited
from the star product. Note by the way that a consistent
description of this Moyal momentum algebra and its application in
integrable models and $2d$-conformal field theories is presented
in our previous works \cite{ref 25, ref 26}.\\\\ In the same lines
of our previous contributions, we try in the present work  to
provide new insights into integrable models and their associated
Lax representation in the $\theta$-deformed framework. The key
step towards achieving this aim is via the Moyal momentum algebra
that we note as ${\widehat \Sigma}(\theta)$. We will use our
convention notation and the properties of the Moyal momentum
algebra ${\widehat \Sigma}(\theta)$ to study some implications of
this algebra in $\theta$-deformed integrable hierarchies and their
Lax representation. We will concentrate on the noncommutative
$sl_2$ KdV$_\theta$ equation and the $\theta$-version of the
Burgers equation and their Lax representations. The derived
properties may naturally be derived for the more general case
namely the $sl_n$
KdV hierarchy.\\\\
In fact, we actually know that the $sl_n$-momentum Lax operators
play a central role in the study of integrable models and more
particularly in deriving higher conformal spin algebras
($w_{\theta}$-algebras) from the $\theta$-extended Gelfand-Dickey
second Hamiltonian structure \cite{ref 23, ref 25, ref 26}. Since
they are also important in recovering $2d$ conformal field
theories via the Miura transformation, we guess that it is
possible to extend this property, in a natural way, to the
$\theta$-deformation case and consider the $\theta$ analogue of
the well-known $2d$-conformal models namely: the
$sl_{2}$-Liouville field theory and its $sl_n$-Toda extensions and
also the Wess-Zumino-Novikov-Witten conformal model.\\\\
This work is presented as follows: We give in \emph{section 2} our
convention notations with a recall of the basic lines of the Moyal
momentum algebra. \emph{Section 3} is devoted to a set up of the
Lax pair representation of special $\theta$-integrable integrable
systems namely the KdV$_\theta$ and Burgers$_\theta$ systems and
\emph{section 4} is for concluding remarks.
\section{Basic Notions}
This section will be devoted to a set up of the background that
will be used in our analysis. For more details about the origin of
the notation used here we refer to \cite{ref 25, ref 26}.\\\\
{\bf 1}. We first start by recalling that the functions often
involved in the two dimensional phase-space are arbitrary
functions which we generally indicate by $f(x,p)$ where the
variable $x$ stands for the space coordinate
while $p$ describes the momentum coordinate. \\\\
{\bf 2}. With respect to this phase space, we have to precise that
the constants $f_0$ are defined such that
\begin{equation}
 \partial_x f_0=0= \partial_p {f_0}.
\end{equation}\\\\
{\bf 3}. The functions ${u_i}(x,t)$ depending on an infinite set
of variables $t_1 =x, t_2,t_3,...,$ are independent from the
momentum coordinates which means
\begin{equation}
\partial_{p} u_{i} (x,t)=0.
\end{equation}
The index $i$, describes the conformal weight of the field $u_i
(x,t)$. These functions can be considered in the complex language
framework as
being the analytic (conformal) fields of conformal spin $i=1,2,...$.
\\\\
{\bf 4}. Other objects usually used are the ones given by
\begin{equation}
u_i (x,t)\star p^j,
\end{equation}
which are objects of conformal weight $(i+j)$ living on the
non-commutative space parametrized by $\theta$. Through this work,
we will use the following convention notations $[u_i]= i$,
$[\theta] =0$ and  $[p]=[\partial _x]=-[x]=1$,
where the symbol $[\hspace{0,5 cm}]$ stands for the conformal dimension of the used objects.\\
{\bf 5}. The star product law, defining the multiplication of
objects in the non-commutative phase space, is given by the
following expression
\begin{equation}
f(x,p)\star g(x,p)= \sum_{s=0}^{\infty}
\sum_{i=0}^{s}{\frac{\theta ^s}{s!}} (-)^{i} c _{s}^{i}
(\partial_{x}^{i}\partial_{p}^{s-i}f)(\partial_{x}^{s-i}\partial_{p}^{i}g),
\end{equation}
with $c _{s}^{i}=\frac {s!}{i!(s-i)!} $.\\
{\bf 6}. The conventional Moyal bracket is defined as
\begin{equation}
\{f(x,p), g(x,p)\}_ {\theta} =\frac {f \star g - g \star
f}{2\theta},
\end{equation}
where $\theta$ is the noncommutative parameter, which, considered
as a constant in this approach{\footnote{The non constant $\theta$
parameter is shown to be important in the noncommutative geometry
framework related to current subject of string theory and D-brane
physics, see for instance \cite{ref 24}
therein}}.\\\\
{\bf 7}. To distinguish the classical objects from the
$\theta$-deformed ones, we consider the following convention notations \cite{ref 25, ref
26}:\\\\
a)$\widehat{\Sigma} _{m}^{(r,s)}$: This is the space of momentum
differential operators of conformal weight $m$ and degrees $(r,s)$
with $r\leq s$. Typical operators of this space are given by
\begin{equation}
\sum _{i=r}^{s}u_{m-i}\star p^{i}.
\end{equation}\\\\
b)$\widehat{\Sigma} _{m}^{(0,0)}$: This is the space of functions
of conformal weight $m$; $ m\in Z$, which may depend on the
parameter $\theta$. It coincides in the classical limit $\theta
\rightarrow \theta_{l}$\footnote{Usually the standard limit is
taken such that $\theta_{limit}=0$. In the present analysis, the
standard limit is shifted by $\frac{1}{2}$ such that
$\theta_{l}\rightarrow\theta_{limit}+\frac{1}{2}$. Thus taking the
standard limit to be $\theta_{limit}=0$ is equivalent to set
$\theta_{l}=\frac{1}{2}$. The origin of this shift belongs to the
consistent $W_{\theta}^{3}$-Zamolodchikov algebra construction
\cite{ref 25, ref 26}}, with the ring of analytic fields involved
into the construction of conformal symmetry and $W$-extensions.\\\\
c)${\widehat\Sigma}_{m}^{(k,k)}$: Is the space of momentum
operators type,
\begin{equation}
u_{m-k}\star p^{k}.
\end{equation}
d)$\theta$-\textbf{Residue operation:} $\widehat{Res}$
\begin{equation}
\widehat{Res}(\alpha \star p^{-1})=\alpha.
\end{equation}
{\bf 8. The Moyal Momentum algebra} \\\\
We denote this algebra in our convention notation by ${\widehat
\Sigma} (\theta)$. This is the algebra based on arbitrary momentum
differential operators of arbitrary conformal weight $m$ and
arbitrary degrees $(r,s)$. Its obtained by summing over all the
allowed values of spin (conformal weight) and degrees in the
following way:
\begin{equation}
{\widehat \Sigma} (\theta) = \oplus _{r\leq s} \oplus _{m \in
Z}{\widehat \Sigma}_{m}^{(r,s)}.
\end{equation}
${\widehat \Sigma} (\theta)$ is an infinite dimensional momentum
algebra which is closed under the Moyal bracket without any
condition. A remarkable property of this space is the possibility
to introduce six infinite dimensional classes of momentum
sub-algebras related to each other by special duality relations.
These classes of algebras are given by ${\widehat
\Sigma}_{s}^{\pm},$ with $s=0,+,-$ describing respectively the
different values of the conformal spin which can be zero, positive
or negative. The $\pm$ upper indices stand for the values of the
degrees quantum numbers, for more details see \cite{ref 7, ref 25,
ref 26}.\\\\
{\bf 9. Algebraic structure of The space ${\widehat \Sigma}
_{m}^{(r,s)}$}:\\\\ To start, let's precise that this space
contains momentum operators of fixed conformal spin m and degrees
(r,s) of type
\begin{equation}
{\mathcal {L}}_{m}^{(r,s)}(u)=\sum _{i=r}^{s}u_{m-i}(x)\star p^i.
\end{equation}
These are $\theta$-differentials whose operator character is
inherited from the star product law defined as in eq.(6). Using
this relation, it is now important to precise how the momentum
operators act on arbitrary functions $f(x,p)$ via the star
product.\\\\
Performing computations based on eq.(6), we find the following
$\theta$- Leibnitz rules:
\begin{equation}
p^{n} \star f(x,p) = \sum _{s=0}^{n} \theta ^{s} c_{n}^{s} f^{(s)}(x,p) p^{n-s},
\end{equation}
and
\begin{equation}
p^{-n} \star f(x,p) = \sum _{s=0}^{\infty} (-)^{s} \theta ^{s}
c_{n+s-1}^{s} f^{(s)}(x,p) p^{-n-s}.
\end{equation}
where $f^{(s)}=\partial_{x}^{s}f$ is the prime derivative.
We also find the following expressions for the Moyal bracket:
\begin{equation}
\begin{array}{lcl}
\{p^n, f\}_{\theta} &=&\sum _{s=0}^{n} \theta ^{s-1} c_{n}^{s}\{\frac {1-(-)^s}{2}\} f^{s} p^{n-s},\\\\
\{p^{-n}, f\}_{\theta} &=& \sum _{s=0}^{\infty} \theta ^{s-1} c_{s+n-1}^{s}\{\frac {(-)^{s}-1}{2}\} f^{s} p^{-n-s},
\end{array}
\end{equation}
These equations don't contribute for even values of $s$ as we can show in the following few examples
\begin{equation}
\begin{array}{lcl}
\{p, f\}_{\theta} &=&f'\\\\
\{p^2, f\}_{\theta} &=&2f'p\\\\
\{p^3, f\}_{\theta} &=&3f'p^2+{\theta}^2f'''
\end{array}
\end{equation}
and
\begin{equation}
\begin{array}{lcl}
\{p^{-1}, f\}_{\theta} &=&-f'p^{-2}-{\theta}^{2}f'''p^{-4}-...- {\theta}^{2k}f^{(2k+1)}p^{-2k-2}-...\\\\
\{p^{-2}, f\}_{\theta} &=&-2f'p^{-3}-4{\theta}^{2}f'''p^{-5}-...- (2k+2){\theta}^{2k}f^{(2k+1)}p^{-2k-3}-...\\\\
\{p^{-3}, f\}_{\theta} &=&-3f'p^{-4}-10{\theta}^{2}f'''p^{-6}-...- \frac{(2k+3)(2k+2)}{2}{\theta}^{2k}f^{(2k+1)}p^{-2k-4}-...
\end{array}
\end{equation}
Special Moyal brackets are given by
\begin{equation}
\begin{array}{lcl}
\{p, x\}_{\theta} &=&1\\\\
\{p^{-1}, x\}_{\theta} &=&-p^{-2}
\end{array}
\end{equation}
Now, having derived and discussed some important properties of the Leibnitz rules,
we can also remark that the momentum operators $p^i$ satisfy the algebra
\begin{equation}
p^n \star p^m =p^{n+m},
\end{equation}
which ensures the suspected rule
\begin{equation}
\begin{array}{lcl}
p^{n} \star (p^{-n}\star f) &=& f\\\\
(f\star p^{-n} )\star p^{n}&=&f.
\end{array}
\end{equation}
\section{The $\theta$-deformation of the Lax representation}
The aim of this section is to present some results related to the
Lax representation of Moyal $\theta$-integrable hierarchies. Using
the convention notations and the analysis presented previously and
developed in \cite{ref 25, ref 26}, we perform consistent
algebraic computations, based on the Moyal-momentum analysis, to
derive explicit Lax pair operators of some integrable systems in
the $\theta$-deformation framework.\\\\
We underline that the present formulation is based on the (pseudo)
momentum operators $p^n$ ad $p^{-n}$ instead of the (pseudo)
operators $\partial^n$ and $\partial^{-n}$ used in several works.
We note also that the obtained results are shown to be compatible
with the ones already established in literature \cite{ref 27}.\\\\
Note by the way that the notion of integrability of the concerned
nonlinear differential equations is defined in the sense that
these equations may be linearizable.\\\\
To start, let's recall that the $sl_n$-Moyal KdV hierarchy is
defined as
\begin{equation}
\frac{\partial {\mathcal L}}{\partial t_{k}}=\{({\mathcal
L}^{\frac{k}{2}})_{+} ,{\mathcal L}\}_{\theta}.
\end{equation}
Explicit computations related to these hierarchies are presented
in \cite{ref 25, ref 26}. Working these hierarchies, we was able
to derive, among others, for the $sl_2$ case up to the flow $t_9$,
the following KdV-hierarchy equations
\begin{equation}
\begin{array}{lcl}
{u}_{t_{1}}&=&{u'},\\\\
{u}_{t_{3}}&=&\frac{3}{2}uu'+{\theta^2}u''',\\\\
{u}_{t_{5}}&=&\frac{15}{8}u^{2}u'+5{\theta^2}(u'u''+\frac{1}{2}uu''')+{\theta}^{4}u^{(5)},\\\\
{u}_{t_{7}}&=&{\frac{35}{16}} u^{3}u^{\prime }+\frac{35}{8}{\theta }%
^{2} ( {4} uu^{\prime }u^{\prime \prime }+ u^{\prime 3}+
u^{2}u^{\prime \prime \prime } ) +\frac{7}{2}( uu^{(5)}+3
u^{\prime }u^{(4)}+{5} u^{\prime \prime }u^{\prime \prime \prime }
) {\theta }^{4}+{\theta }^{6}u^{(7)}, \\\\
...
\end{array}
\end{equation}
Actually this construction which works well for the $sl_2$-KdV
hierarchy is generalizable to higher order KdV hierarchies, namely
the $sl_n$-KdV hierarchies.\\\\
The basic idea of the Lax formulation consists first in
considering a noncommutative integrable system which possesses the
Lax representation such that the following $\theta$-deformation
Moyal bracket
\begin{equation}
\{L, T+\partial_t\}_{\theta}=0,
\end{equation}
is equivalent to the $\theta$-differential equation that we
consider from the beginning and that is nonlinear in general with $\partial_{t}\equiv \frac{\partial}{\partial t}$.\\
Equation(23) and the associated pair of operators $(L,T)$ are
called the Lax differential equation and the Lax pair,
respectively. The differential operator $L$ defines the integrable
system which we should fix from the beginning.\\\\
Note that the way with which ones to writes the Lax equation as in
(23) is equivalent to that in (21) namely
\begin{equation}
\{L, T+\partial_t\}_{\theta}\equiv\{L, ({\mathcal
L}^{\frac{k}{2}})_{+}+\partial_{t_k}\}_{\theta}=0,
\end{equation}
where the operator $T$ is the analogue of $({\mathcal
L}^{\frac{k}{2}})_{+}$ describing then an operator of conformal
spin $k$.\\\\
This equation, written in terms of the function $u(x,t)$, is in
general a non linear differential equation belonging to the ring
${\widehat{\Sigma} _{k+2}^{(0,0)}}$. In the present case of
$sl_2$-KdV systems we have $k=3$.\\\\
As it's shown in  \cite{ref 27}, the meaning of Lax
representations in $\theta$-deformed spaces would be vague.
However, they actually have close connections with the bi-complex
method \cite{ref 28} leading to infinite number of conserved
quantities, and the (anti)-self-dual Yang-Mills equation which is
integrable in the context of twistor descriptions and ADHM
constructions \cite{ref 29, ref 30}.
\\\\ Now, let us apply the $\theta$-deformation Lax-pair generating technique.
Usually, it's a method to find a corresponding $T$-operator for a
given $L$-operator. Finding the operator $T$ satisfying (23) is
not an easy job in the general case. For this reason, one have to
make
some constraints on the operator $T$ namely:\\\\
{\bf Ansatz for the operator $T$}:
\begin{equation}
T=p^n\star L^{m}+ T',
\end{equation}
where $p^n$ are momentum operators acting on arbitrary function
$f(x,p)$ as shown in \emph{section 2}. Note by the way that the
notation $T'$ have nothing to do with the prime derivative. With the
previous ansatz, the problem reduces to that for the $T'$-operator
which is determined by hand so that the Lax equation should be a
differential equation
belongings to the ring ${\widehat{\Sigma}^{(0,0)}}$.\\
The best way to understand what happens for the general case, is
to focus on the following examples:\\
{\bf Example 1: The $\theta$-deformation of the KdV system.}\\\\
The $L$-operator for the noncommutative KdV equation is given, in
the momentum space configuration, by
\begin{equation}
L=p^2+u(x,t),
\end{equation}
with
\begin{equation}
L\in {\widehat{\Sigma} _{2}^{(0,2)}}/{\widehat{\Sigma}
_{2}^{(1,1)}},
\end{equation}
where ${\widehat{\Sigma} _{2}^{(1,1)}}$ is the one dimensional
subspace generated by objects of type $\xi_{1}(x,t)\star p$ and
$\xi_{1}(x,t)$ is an arbitrary function of conformal spin $1$.\\\\
Reduced to $n=1=m$, for the deformed $sl_2$ KdV$_\theta$ system,
the ansatz (25) can be written as follows\footnote{One can also
introduce the following definition: $T\equiv (p\star L)_{s}+ T'$,
with the convention $(p\star L)_{s}\equiv \frac{(p\star L+L\star
p)}{2}$ describing the symmetrized part of the operator $p\star
L$}
\begin{equation}
T=p\star L+ T'.
\end{equation}
The operator $T$ in this case ($k=3$), is shown to behaves as
$({\mathcal L}^{\frac{3}{2}})_{+}$ with $\partial_{t_3}\equiv
\frac{\partial}{\partial t_3}$.\\\\
Simply algebraic computations give
\begin{equation}
\{L, T'\}_{\theta}=u'p^2-2\theta
u''p+\frac{\dot{u}}{2\theta}+(uu'+\theta^2u'''),
\end{equation}
Next, our goal is to be able to extract the Lax differential
equation, namely, the KdV$_\theta$ equation. Before that, we have
to make a projection of the operator $\{L, T'\}_{\theta}$ on the
ring ${\widehat{\Sigma} _{3+2}^{(0,0)}}$. This projection is
equivalent to cancel the effect of the terms of momentum in (29),
namely the term $u'\textbf{p}^2$ and $2\theta
u''\textbf{p}$. To do that, we have to consider the following property: \\\\
{\bf Ansatz for $T'$}:\\
\begin{equation}
T'=X\star p+ Y,
\end{equation}
where $X$ and $Y$ are arbitrary functions on $u$ and its
derivatives.
\\\\
Next, performing straightforward computations, with $T'=Xp-\theta
X'+Y$ lead to
\begin{equation}
\{L,T'\}_{\theta}=2X'p^{\textbf{2}}+(\{u,X\}_{\theta}-2\theta
 X''+2Y')p+ (-Xu'-\theta \{u,X'\}+\{u,Y\} )
\end{equation}
Identifying (29) and (31), leads to the following constrains on
the parameters $X$ and $Y$
\begin{equation}
X=\frac{1}{2}u_{2}+a,
\end{equation}
\begin{equation}
Y=-\frac{1}{2}\theta u_{2}'+b,
\end{equation}
with the following nonlinear differential equation
\begin{equation}
-\frac{\dot{u}}{2\theta}=\frac{3}{2}uu'+\theta^{2}u'''.
\end{equation}
where the constants $a$ and $b$ are to be omitted for a matter of
simplicity. The last equation is noting but the $sl_2$
KdV$_\theta$ equation. This deformed equation contains also a non
linear term $\frac{3}{2}uu'$.
\\\\We have to
underline that the $sl_2$ KdV$_\theta$ equation obtained through
this Lax method belongs to the same class of the KdV equation
derived in \cite{ref 25, ref 26} namely
\begin{equation}
\dot{u}=\frac{3}{2}uu'+\theta^{2}u'''.
\end{equation}
In fact, performing the following scaling transformation
$\partial_{t_{3}}\rightarrow -2\theta\partial_{t_3}$ we recover
exactly (35). The term $\frac{1}{2\theta}$ appearing in (34) as
been the coefficient of the evolution part $\dot{u_{2}}$ of the NC
KdV equation can be simply shifted to one due to consistency with
respect to the classical limit $\theta_{l} \sim \frac{1}{2}$.
\\\\ To summarize, the momentum Lax pair operators, associated to the
deformed $sl_2$-KdV system, are explicitly given by
\begin{equation}
L_{KdV}=p^{2}+u_{2}(x,t),
\end{equation}
and
\begin{equation}
T_{KdV}=p^3+\frac{3}{2}p\star u_{2}(x,t)-\frac{3}{2}\theta
u_{2}'(x,t);
\end{equation}
with $T'=\frac{1}{2}\textbf{p}\star u_{2}(x,t)-\frac{3}{2}\theta u'_{2}(x,t)$\\\\
Note that, the same results can obtained by using the Gelfand-Dickey
(GD) formulation based on formal (pseud) differential operators
$\partial ^{\pm n}$ instead of the Moyal momentum
ones.\\\\
This first example shows, among others, the consistency of the Moyal
momentum formulation in describing integrable systems and the
associated Lax pair generating technics in the same way as the
successful GD formulation \cite{ref 7, ref 8}.\\\\
{\bf Example 2: The Burgers$_\theta$ Equation}\\
Let us apply the same $\theta$-deformation Lax technics, presented
presented previously, to derive the $\theta$-deformation of the
Burgers equation. Actually, our interest in this equation comes
from the several important properties that are exhibited in the
standard case (commutative). Before going into applying the
$\theta$-deformation technics let's first recall some few known
properties of the standard Burgers
equation.\\\\
{\bf P1}: The Burgers equation is defined on the $(1+1)$-
dimensional space time. In the standard pseudo-differential
operator formalism, this equation is associated to the following
L-operator
\begin{equation}
L_{Burgers}=\partial_x+u_{1}(x,t)
\end{equation}
where the function $u_1$ is of conformal spin one. Using our
convention notations, we can set $L\in {\widehat
\Sigma}^{(0,1)}_{1}$.\\
{\bf P2}: With respect to the previous L-operator, the non linear
differential equation of the Burgers equation is given by
\begin{equation}
\dot{u}_{1}+\alpha u_{1}u'_{1}+\beta u''_{1}=0,
\end{equation}
where $\dot{u}=\frac{\partial u}{\partial t}$ and
$u'=\frac{\partial u}{\partial x}$. The dimensions of the
underlying objects are given by $[t]=-2=-[\partial_t]$, $[x]=-1$
and $[u]=1$.\\\\
{\bf P3}: On the commutative space-time, the Burgers equation can
be derived from the Navier-Stokes equation and describes real
phenomena, such as the turbulence and shock waves. In this sense,
the Burgers equation draws much attention amongst many integrable
equations.\\\\
{\bf P4}: It can be linearized by the Cole-Hopf transformation
\cite{ref 31}. The linearized equation is the diffusion equation and
can be solved by Fourier transformation for given boundary
conditions.\\\\
{\bf P5}: The Burgers equation is completely integrable \cite{ref 32}.\\\\
Now, we are ready to look for the $\theta$-deformation version of
the Burgers equation. For that, we consider the $L$-operator of
this equation in the Moyal momentum language, namely
\begin{equation}
L=p+u_{1}(x,t).
\end{equation}
dealing, as noticed before, to the space ${\widehat
\Sigma}^{(0,1)}_{1}$. This is a local differential operator of the
generalized $n$-KdV hierarchy's family $(n=1)$, obtained by a
$\theta$-truncation of a pseudo momentum operator of KP hierarchy
type
\begin{equation}
L=p+u_{1}(x,t)+u_{2}(x,t)\star p^{-1}+u_{3}(x,t)\star p^{-2}+...,
\end{equation}
of the space ${\widehat \Sigma}^{(-\infty,1)}_{1}$. The local
truncation is simply given by
\begin{equation}
{\widehat \Sigma}^{(-\infty,1)}_{1}\rightarrow {\widehat
\Sigma}^{(0,1)}_{1}\equiv [{\widehat
\Sigma}^{(-\infty,1)}_{1}]_{+}\equiv {\widehat
\Sigma}^{(-\infty,1)}_{1}/{\widehat \Sigma}^{(-\infty,-1)}_{1},
\end{equation}
or equivalently
\begin{equation}
L_{1}(u_i)=p+\Sigma _{i=0}^{\infty}u_{i}\star p^{1-i} \rightarrow
p+u_{1}\equiv[L_{1}(u_i)]_{+},
\end{equation}
where the symbol $(X)_{+}$ defines the local part (only positive
powers of $p$) of a given pseudo operator $X$.
\\\\
The $\theta$-deformation of the Burgers equation is said to have
the Lax representation if there exists a suitable pair of
operators $(L,T)$ so that the Lax equation
\begin{equation}
\{p+u_1, T+\partial_t\}_{\theta}=0,
\end{equation}
reproduces the $\theta$-deformation version of the Burgers non
linear differential equation. Following the same steps developed
previously for the $sl_2$ KdV$_\theta$ systems, we consider the
following ansatz for the operator $T$:
\begin{equation}
T=p\star L+ T',
\end{equation}
or
\begin{equation}
T=p^2+ u_{1}p+ \theta u'_{1}+ T'.
\end{equation}
Then, performing straightforward computations, the
Burgers$(theta)$ Lax equation (44) reduces to
\begin{equation}
\{p+u,T'\}_{\theta}=u'p+(uu'-\theta u''+\frac{\dot{u}}{2\theta})
\end{equation}
Next, one have also the go through a constraint equation for the operator $T'$, namely the
\\\\
{\bf Ansatz for $T'$}:
\begin{equation}
T'=A\star p+ B,
\end{equation}
where $A$ and $B$ are arbitrary functions for the moment.With this
new ansatz for $T'$, we have
\begin{equation}
\{p+u,T'\}_{\theta}=(A'+\{u,A\}_{\theta})p+(-Au'-\theta A''-\theta
\{u,A'\}_{\theta}+B'+\{u,B\}_{\theta})
\end{equation}
\\Identifying (47) and (49) leads to the following constraints
equations
\begin{equation}
u'=A'+\{u,A\}_{\theta})
\end{equation}
and
\begin{equation}
(u+A)u'+\frac{\dot{u}}{2\theta}=B'+\{u,B\}_{\theta}+\theta
\{A',u\}+\theta(u''-A'')
\end{equation}
A natural solution of the first constraint equation (50) is $A=u$.
This leads to a reduction of (51) to
\begin{equation}
2uu'+\frac{\dot{u}}{2\theta}=B'+\{u,B\}
\end{equation}
Actually this is the $\theta$-deformationof the Burgers equation,
which is also the projection of the Lax equation (44) to the ring
of vanishing degrees in momenta namely the space
${\widehat{\Sigma}
_{1}^{(0,0)}}$.\\\\
Since $\{u,\partial_t\}_{\theta}=-\frac{\dot u}{2\theta}$, a non
trivial solution of the parameter $B$ in equation (52) consists in
setting $B\equiv u^2-\frac{\partial}{\partial t}$. But, since this
nontrivial solution of $B$ masks the Burgers$(\theta)$
equation, it's a non desirable thing.\\\\
Remarking also that $[B]=2$, we use this dimensional arguments and
set
\begin{equation}
B=\xi u'+\eta u^2
\end{equation}
with $\xi$ and $\eta$ are arbitrary coefficient numbers. Injecting
this expression into (52) gives the final expression of the
$\theta$-deformation of the Burgers equation namely
\begin{equation}
\frac{\dot{u}}{2\theta}+2(1-\eta)uu'-\xi u''=0
\end{equation}
whose Lax pair in the Moyal momentum formalism are explicitly
given by
\begin{equation}
L_{Burgers}=p+u_{1}(x,t)
\end{equation}
and
\begin{equation}
T_{Burgers}=p^2+2u_{1}(x,t)p+\eta u_{1}^2(x,t)+\xi u_{1}'(x,t)
\end{equation}
\newpage
\section{Concluding Remarks}
To summarize the main lines of the present work, let's present some important remarks:\\
{\bf 1.} We have presented a systematic study of the generating
Lax pair operators technics in the Moyal momentum framework. The
essential results deals with the derivation of the
$\theta$-deformation of the KdV and Burgers systems.\\
{\bf 2.} Concerning the $\theta$-deformed derived KdV system, this
is an integrable model due to the existence of a $\theta$-deformed
Lax pair operators $(L, T)$. This existence is an important
indication of integrability, but we guess that the integrability
of this model is the underlying conformal symmetry,
shown to play a similar role as in the standard commutative case.\\
{\bf 3.} Concerning the $\theta$ parameter appearing in the
KdV$_\theta$ equation, we already mentioned before that its
classical limit is not the same as the standard one corresponding
usually in taking $\theta_{limit}=0$. This is because the KdV
hierarchy systems, in general, deal with the conformal symmetry
and its $W_s$-extensions describing non trivial Lie algebras.
Recall that in the language of $2d$-conformal field theory, the
above mentioned currents $W_{s}$ are taken in general as primary
satisfying the OPE \cite{ref 2}
\begin{equation}
T(z)W_{s}(\omega )=\frac{s}{(z-\omega )^{2}}W_{s}(\omega )+\frac{%
W_{s}^{\prime }(\omega )}{(z-\omega )}.
\end{equation}
As we are interested in the $\theta$-deformation case, we have to
add that the spaces $\Sigma ^{(0,0)}_{k}$ are $\theta$-depending
and the corresponding $W_{\theta}$-algebra is shown to exhibit new
properties related to the $\theta$ parameter and reduces to the
standard $W$-algebra once some special limits on the $\theta$
parameters are performed.\\
As an example, consider for instance the $W_{\theta}^{3}$-algebra
\cite{ref 23, ref 25} generalizing the Zamolodchikov algebra
\cite{ref 8, ref 9}. The conserved currents of this extended algebra
are shown to take the following form
\begin{equation}
\begin{array}{lcl}
w_2 & = & u_{2}\\
w_3 & = & u_{3}-\theta u_{2}^{^{\prime}}
\end{array}
\end{equation}
which coincides with the standard case once $\theta =\frac{1}{2}$.\\
{\bf 4.} Concerning the $\theta$-deformation of the Burgers
equation (54) that we consider in the second example, it's also an
integrable equation whose Lax pair operators are explicitly
derived (55-56). Note for instance that the Burgers Lax operator
$L_{Burgers}=p+u_{1}(x,t)$ is a momentum operator belongings to
the space
${\widehat\Sigma}_{1}^{(0,1)}$. \\
 {\bf 5.} As a first checking of integrability for the Burgers$(\theta)$ system, we
 proceeded to an explicit derivation of the Lax pair operators
 $(L,T)$ giving the following differential equation
$2uu'+\frac{\dot{u}}{2\theta}=B'+[u,B]$. The idea is to solve this
equation in terms of the coefficient parameter $B$ such that it can
reduces to the non linear Burgers equation belonging to the space
${\widehat\Sigma}_{3}^{(0,0)}$. Solving this equation give
explicitly the requested Lax operator.
Presently, we have two possible solutions:\\
{\bf The first solution}:
\begin{equation}
B_1=u^{2}-\partial_t,
\end{equation}
defines the parameter $B$ in terms of time derivative and it's not
suitable for us because it masks the nonlinear differential equation
that we are looking for. \\
{\bf The second solution}:
\begin{equation}
B_2=\xi u'+\eta u^{2}
\end{equation}
which is consistent with dimensional arguments maps us to the
$\theta$-deformation of the Burgers equation (54).\\
{\bf 6.} We should also underline that the importance of this
study comes also from the fact that the results obtained in the
framework of Moyal momentum are similar to those coming by
using the Gelfand-Dickey pseudo operators approach.\\
{\bf 7.} Finally, a significant question is to know if there is a
possibility to establish a correspondence between the two Systems.
The reason is that for the $\theta$-deformation of the KdV model,
the problem of integrability does not arise in the same way as
it's for the Burgers$(\theta)$ equation.\\
The first one is mapped to conformal field theory through the
Liouville model. This is in fact a strong indication of
integrability in favor of KdV$_\theta$ equation which could
help more to understand the $\theta$-deformation of the Burgers system.\\
Such a correspondence between the two systems, if it can be
realized, might bring new insights towards understanding much more
their integrability. This and others questions will be done in a
future work.\\\\
{\bf Acknowledgements} \\
We would like to thank the Abdus Salam International Center for
Theoretical Physics (ICTP) for hospitality and acknowledge the
considerable help of the High Energy Section and its head S.
Randjbar-Daemi. This work was done within the framework of the
Associateship Scheme of the Abdus Salam International Center for
Theoretical Physics, Trieste, Italy. We would like to thank the
Associateship office and more particularly its staff for the
quality of service and for the permanent availability. A.E. and
M.B.S present special acknowledgements to OEA-ICTP and to its head
George Thompson for valuable scientific helps in the context of
Network. They also wish to thank the University Ibn Tofail, the
Faculty of Sciences at Kenitra and the protas III programm D12/25
CNR, Morocco. D.O present acknowledgements to The Sweden
International Development Cooperation Agency(SIDA)) for financial
supports.

\end{document}